\documentstyle[12pt]{article}
\newcommand\bea{\begin{eqnarray}}
\newcommand\eea{\end{eqnarray}}
\begin{document}
\thispagestyle{empty}
\bibliographystyle{unsrt}
\setlength{\baselineskip}{18pt}
\parindent 24pt
\vspace{60pt}

\hfill{IFA-FT-396-1994}
\begin{center}{
{\huge
Damped quantum harmonic oscillator:\\
density operator and related quantities}\\[7mm]
\vskip 1truecm
A. Isar\\
{\it Department of Theoretical Physics, Institute of Atomic Physics\\
POB MG-6, Bucharest-Magurele, Romania\\
Internet: {\rm isar@ifa.ro}}
}\end{center}
\vskip 1truecm
\begin{abstract}
\vskip 0.5truecm
A closed expression for the density operator
of the damped
harmonic oscillator
is extracted
from the master equation based on the Lindblad theory for open quantum systems.
The entropy and effective temperature of the system are
subsequently calculated
and their temporal behaviour is surveyed by
showing how these quantities relax to their equilibrium values.
The entropy for a state characterized by a Wigner distribution function
which is Gaussian in form is found to depend only on the variance of the
distribution function.

\end{abstract}
\vskip 3truecm
Shortened title: Damped quantum harmonic oscillator

\newpage
\setcounter{page}{1}
\section{Introduction}

In the last two decades, the problem of dissipation in quantum mechanics, i.e.
the consistent description of open quantum systems, was investigated by various
authors [1-5]  (for a recent review
on open quantum systems
see ref. \cite{rev}).
It is commonly understood \cite{d2,d1} that dissipation
in an open system results from microscopic reversible interactions between
the observable system and the environment.
Because dissipative processes imply irreversibility and,
therefore, a preferred direction in time, it is generally thought that quantum
dynamical semigroups are the basic tools to introduce dissipation in quantum
mechanics. In the Markov approximation
the most general form of the generators of such semigroups was given
by Lindblad \cite{l1}. This formalism has been studied for the case of damped
harmonic oscillators \cite{l2,ss} and applied to various physical phenomena,
for
instance, the damping of collective modes in deep inelastic collisions in
nuclear physics \cite{i1}.
In \cite{i2} the Lindblad master
equation for the harmonic oscillator
was transformed into Fokker-Planck equations for quasiprobability distributions
and a comparative study was made for the Glauber $P$, antinormal ordering $Q$
and Wigner $W$ representations.
In \cite{a} the density matrix for the coherent
state representation and the Wigner distribution function subject to different
types of initial conditions were obtained for the damped harmonic oscillator.
A remarkable feature resulting from the preceding analysis \cite{ass} is the
agreement in form of the Lindblad master and Fokker-Planck equations
with the corresponding equations familiar from quantum optics. Quantum
optics is generally formulated with the help of the quantum mechanics
of the damped harmonic oscillator \cite{d2,d1} and the corresponding
Brownian motion.
Generally, the equation of motion treated in the theory of Brownian
motion is the master equation satisfied by the density operator or the
Fokker-Planck equation for the distribution function. The master and
Fokker-Planck equations related to the problem of Brownian motion have
been fully reviewed in the past \cite{d2,d1,hh,hr}.

In the present study we are also concerned
with the observable system of a harmonic oscillator which interacts with
an environment.
The aim of this work is to explore further the physical aspects of the
Fokker-Planck
equation which is the $c$-number equivalent equation to the master
equation. Generally the master equation for the density operator gains
considerably in clarity if it is represented in terms of the Wigner
distribution function which satisfies the Fokker-Planck equation.
We shall describe,
within the
Lindblad theory for open quantum systems,
how
the system under consideration evolves to a final state of equilibrium
by calculating first an explicit form of the density operator
satisfying the operator master equation based on the Lindblad dynamics.
We subsequently derive the entropy and the effective temperature of the
quantum-mechanical system in a state characterized by a Wigner
distribution function which is Gaussian in form and show how these quantities
relax to their equilibrium values.

Entropy is a quantity which may be interpreted physically as a measure
of the lack of knowledge of the system. The idea of effective
temperature associated with the Bose occupation distribution has
already been introduced \cite{lw} in relation to the entropy which has been
derived in the framework of quantum theory of harmonic oscillator
relaxation. The derivation of the entropy associated with an infinite
coupled harmonic oscillator chain has similarly been elaborated for
classical \cite{rh} and quantum mechanical systems \cite{a1} represented by a
phase space distribution function.
In the present paper we derive first the density
operator of the damped harmonic oscillator in the Lindblad theory by
applying a procedure similar to the techniques developed in the description
of quantum relaxation \cite{lw,a1,m,j1}.
When we denote by $ \rho(t)$ the density operator of the damped harmonic
oscillator in the Schr\"odinger
picture, the entropy  $S(t) $ is given by
\bea   S(t)= -k{\rm Tr}(\rho\ln\rho)=
-k\sum_{m,n}<m\vert\rho\vert n><n\vert\ln\rho\vert m>,     \eea
where $k$ is Boltzmann's constant. While the explicit form of the general
density matrix $ <m\vert\rho\vert n> $ is now available \cite{ass}, the
evaluation of
the matrix element of the logarithmic operator $\ln\rho$ is not an easy
task. An alternative way of calculating the entropy is to compute straightway
the expectation value of the logarithmic operator
$<\ln\rho>={\rm Tr}(\rho\ln\rho).$
Accordingly, the problem amounts to derive
the explicit form of the density operator.

The content of this paper is arranged as follows. In Sec. 2 we write
the master equation for the density operator of the harmonic
oscillator. Sec. 3 derives an explicit form of the density operator involved
in
the
Lindblad master equation.
Sec. 4 formulates the entropy and time-dependent temperature using the
explicit form of the density operator and discusses their temporal
behaviour.
Finally, concluding remarks are given in Sec. 5.

\section{Master equation for the damped quantum harmonic oscillator}

The rigorous formulation for introducing the dissipation into a quantum
mechanical system is that of quantum dynamical semigroups \cite{d,s,l1}.
According to
the axiomatic theory of Lindblad \cite{l1}, the usual von Neumann-Liouville
equation ruling the time evolution of closed quantum systems is replaced in the
case of open systems by the following equation for the density operator $\rho$:
\bea   {d\Phi_{t}(\rho)\over dt}=L(\Phi_{t}(\rho)).     \eea
Here, $\Phi_{t}$ denotes the dynamical semigroup describing the irreversible
time evolution of the open system in the Schr\"odinger representation and $L$
the infinitesimal generator of the dynamical semigroup $\Phi_t$. Using the
structural theorem of Lindblad \cite{l1} which gives the most general form of
the
bounded, completely dissipative Liouville operator $L$, we obtain the explicit
form of the most general time-homogeneous quantum mechanical Markovian master
equation:
\bea   {d\rho(t)\over dt}=L(\rho(t)),     \eea
where
\bea   L(\rho(t))=-{i\over\hbar}[H,\rho(t)]+{1\over 2\hbar}\sum_{j}([V_{j}
\rho(t),V_{j}^\dagger]+[V_{j},\rho(t)V_{j}^\dagger]).     \eea
Here $H$ is the Hamiltonian of the system. The operators $V_{j}$ and $V_{j}^
\dagger$
are bounded operators on the Hilbert space $\cal H$ of the Hamiltonian.

We should like to mention that the Markovian master equations found in the
literature are of this form after some rearrangement of terms, even for
unbounded Liouville operators. In this connection we assume that the general
form of the master equation given by (3), (4) is also valid for unbounded
Liouville operators.

In this paper we impose a simple condition to the operators $H,V_{j},V_{j}^
\dagger$
that they are functions of the basic observables $\hat q$ and $\hat p$ of the
one-dimensional quantum mechanical system (with $[\hat q,\hat p]=i\hbar
I,$ where $I$ is the identity operator on $\cal H$) of
such kind
that the obtained model is exactly solvable. A precise version for this last
condition is that linear spaces spanned by first degree (respectively second
degree) noncommutative polynomials in $\hat q$ and $\hat p $ are invariant to
the action
of the completely dissipative mapping $L$. This condition implies \cite{l2}
that $V_
{j}$ are at most first degree polynomials in $\hat q$ and $\hat p $ and $H$ is
at most a
second degree polynomial in $\hat q$ and $\hat p $. Then the harmonic
oscillator
Hamiltonian $H$ is chosen of the form
\bea   H=H_{0}+{\mu \over 2}(\hat q\hat p+\hat p\hat q),~~~H_{0}={1\over 2m}
\hat
p^2
+{m\omega^2\over 2}\hat q^2.     \eea
With these choices the Markovian master equation can be written \cite{ss}:
\bea   {d\rho \over dt}=-{i\over \hbar}[H_{0},\rho]-{i\over 2\hbar}(\lambda +
\mu)
[\hat q,\rho \hat p+\hat p\rho]+{i\over 2\hbar}(\lambda -\mu)[\hat p,\rho
\hat q+\hat q\rho]  \nonumber\\
  -{D_{pp}\over {\hbar}^2}[\hat q,[\hat q,\rho]]-{D_{qq}\over {\hbar}^2}
[\hat p,[\hat p,\rho]]+{D_{pq}\over {\hbar}^2}([\hat q,[\hat p,\rho]]+[\hat p,
[\hat q,\rho]]),      \eea
where $D_{pp},D_{qq}$ and $D_{pq}$ are the diffusion coefficients and $\lambda$
the friction constant. They satisfy the following fundamental constraints
\cite{ss}:
\bea   {\rm i})~D_{pp}>0,~~{\rm ii})~D_{qq}>0,~~{\rm iii})~D_{pp}D_{qq}-D_{pq}
^2
\ge
{\lambda}^2{\hbar}^2/4.    \eea
In the
particular case when the asymptotic state is a Gibbs state
\bea   \rho_G(\infty)=e^{-{H_0\over kT}}/{\rm Tr}e^{-{H_0\over kT}},    \eea
these coefficients reduce to
\bea   D_{pp}={\lambda+\mu\over 2}\hbar m\omega\coth{\hbar\omega\over 2kT},
{}~~D_{qq}={\lambda-\mu\over 2}{\hbar\over m\omega}\coth{\hbar\omega\over 2kT},
{}~~D_{pq}=0,    \eea
where $T$ is the temperature of the thermal bath.

\section{Evaluation of the density operator}

There are several techniques available to solve equations such as (6).
By introducing
the real variables $x_1, x_2$ corresponding to the
operators $\hat q, \hat p$:
\bea x_1=\sqrt{m\omega\over 2\hbar}q, ~~x_2={1\over\sqrt{2\hbar m\omega}}p,
\eea
in \cite{i2,a} we have transformed the operator form of the master
equation into the following Fokker-Planck equation satisfied by the Wigner
distribution
function $W(x_1, x_2, t):$
\bea   {\partial W\over\partial t}=\sum_{i,j=1,2}A_{ij}{\partial\over
\partial x_i}(x_jW)+{1\over 2}\sum_{i,j=1,2}Q^W_{ij}{\partial^2\over
\partial x_i\partial x_j}W,    \eea
where
\bea   A=\left(\matrix{\lambda-\mu&-\omega\cr
\omega&\lambda+\mu\cr}\right),~~~
Q^W={1\over\hbar}\left(\matrix{m\omega D_{qq}&D_{pq}\cr
D_{pq}&D_{pp}/m\omega\cr}\right).    \eea
Since the drift coefficients are linear in the variables $x_1$ and $x_2$ and
the diffusion coefficients are constant with respect to $x_1$ and $x_2,$
Eq. (11)
describes an Ornstein-Uhlenbeck process \cite{wu}. Following the method
developed
by Wang and Uhlenbeck \cite{wu}, we solved \cite{a}
this Fokker-Planck equation, subject
to either the wave-packet type or the $\delta$-function type of initial
conditions.

When the Fokker-Planck equation is subject to a Gaussian
(wave-packet) type of the initial condition  ($x_{10}$ and $x_{20}$ are the
initial values of $x_1$ and $x_2$ at $t=0,$ respectively)
\bea W_w(x_1, x_2, 0)={1\over\pi\hbar}\exp\{-2[(x_1-x_{10})^2+(x_2-x_{20})^2
]\},
  \eea
the solution is found to be \cite{a}:
\bea W_w(x_1, x_2)={\Omega\over \pi\hbar\omega\sqrt{-B_w}}\exp\{-
{1\over B_w}
[\phi_w(x_1-\bar x_1)^2+\psi_w(x_2-\bar x_2)^2+\chi_w(x_1-\bar x_1)
(x_2-\bar x_2)]\},  \eea
where
\bea   B_w=g_1g_2-{1\over 4}g_3^2,~~g_1=g_2^*={\mu a\over\omega}e^{2\Lambda t}+
{d_1\over \Lambda}
(e^
{2\Lambda t}-1),~~g_3=2[e^{-2\lambda t}+{d_2\over\lambda}(1-e^{-2\lambda
t})],
    \eea
\bea\phi_w=g_1a^{*2}+g_2a^2-g_3,~\psi_w=g_1+g_2-g_3,~\chi_w=2(g_1a^*+g_
2a)-g_3(a+a^*).  \eea
We have put $a=(\mu-i\Omega)/\omega,
{}~\Lambda=-\lambda-i\Omega,$
{}~$d_1=(a^2m\omega D_{qq}+2aD_{pq}+D_{pp}/m\omega)/\hbar,$
{}~$d_2=(m\omega D_{qq}+2\mu D_{pq}/\omega+D_{pp}/m\omega)/\hbar$
and $\Omega^2=\omega^2-\mu^2.$
The functions $\bar x_1$ and $\bar x_2$, which are also oscillating
functions, are given by \cite{a}:
\bea\bar x_1=e^{-\lambda t}[(x_{10}(\cos\Omega t+{\mu\over\Omega}\sin
\Omega t)+x_{20}{\omega\over\Omega}\sin\Omega t],  \eea
\bea\bar x_2=e^{-\lambda t}[(x_{20}(\cos\Omega t-{\mu\over\Omega}\sin
\Omega t)-x_{10}{\omega\over\Omega}\sin\Omega t].  \eea
The solution (14)
of the Fokker-Planck equation (11), subject to the wave-packet type of initial
condition (13) can be written in terms of the coordinate and momentum
($<\hat A>={\rm Tr}(\rho \hat A)$ denotes the expectation value of an
operator $\hat A$)
as \cite{ss}:
\bea   W(q,p)={1\over 2\pi\sqrt{\delta}}\exp\{-{1\over 2\delta}[\phi
(q-<\hat q>)^2+\psi(p-<\hat p>)^2-2\chi(q-<\hat q>)(p-<\hat p>)]\},
    \eea
where
\bea <\hat q>
=e^{-\lambda t}[(\cos\Omega t+{\mu\over\Omega}\sin\Omega t)
<\hat q(0)>+{1\over m\Omega}\sin\Omega t<\hat p(0)>],\eea
\bea <\hat p>
=e^{-\lambda t}[-{m\omega^2\over\Omega}\sin\Omega t<\hat q(0)>+
(\cos\Omega t-{\mu\over\Omega}\sin\Omega t)<\hat p(0)>],    \eea
\bea  \phi\equiv\sigma_{pp}
=<\hat p ^2>-<\hat p >^2=
-{\hbar m\omega^3\over 4\Omega^2}\phi_w,
\eea
\bea  \psi\equiv\sigma_{qq}
=<\hat q ^2>-<\hat q >^2
=-{\hbar\omega\over 4m\Omega^2}\psi_w,
\eea
\bea  \chi\equiv\sigma_{pq}(t)={1\over 2}<\hat q \hat p +\hat p \hat q >-
<\hat q ><\hat p >
={\hbar\omega^2\over 8\Omega^2}\chi_w,~~\delta=\phi\psi-\chi^2.
\eea

To get the explicit expression of the density operator, we use the relation
$\rho=2\pi\hbar {\bf\it N}\{W_s(q,p)\},$ where $W_s$ is the Wigner
distribution
function in the form of standard rule of association and ${\bf\it N}$ is the
normal ordering operator \cite{w,l} which acting on the function $W_s(q,p)$
moves all $p$ to the right of the $q.$ By the standard
rule of association is meant the correspondence $p^mq^n\to \hat q^n\hat p^m$
between functions of two classical variables $(q,p)$ and functions of two
quantum mechanical canonical operators $(\hat q,\hat p).$ The calculation of
the density operator is then reduced to a problem of transformation of the
Wigner distribution function by the ${\bf\it N}$ operator, provided that $W_s$
is
known. A special care is necessary for the ${\bf\it N}$ operation when the
Wigner
function is in the exponential form of a second order polynomial of $q$ and
$p.$ The Wigner distribution function
(19) corresponds however to the form of the
Weyl rule of association \cite{hw}.
This function can be transformed into the form of
standard rule of association \cite{clm} by
\bea   W_s(q,p)=\exp({1\over 2}i\hbar{\partial^2\over\partial p\partial q})
W(q,p).
\eea
Upon performing the operation on the right-hand side, we get the Wigner
distribution function $W_s,$ which has the same form as the original $W$
multiplied by $\hbar$ but with $\chi-i\hbar/2$ in place of
$\chi:$

\bea   W_s(q,p)={1\over 2\pi\sqrt{\xi}}\exp\{-{1\over 2\xi}[\phi
(q-<\hat q>)^2+\psi(p-<\hat p>)^2-2\chi'(q-<\hat q>)(p-<\hat p>)]\},
    \eea
where
\bea  \xi=\phi\psi-\chi'^2,
{}~~ \chi'=\chi-i{\hbar\over 2}.\eea
The normal ordering operation of the Wigner function $W_s$ in
Gaussian form can be carried out by applying McCoy's theorem \cite{w,l}, which
states that
\bea   [J\exp(-i\hbar\gamma)]^{1/2}\exp(\alpha\hat q^2+\beta\hat p^2+\gamma
\hat q\hat p)={\bf\it N}[\exp(Aq^2+Bp^2+Gqp)],\eea
where $\alpha=A/C,~ \beta=B/C,~ C=\sinh\Gamma/\Gamma J, ~\Gamma=-i\hbar(\gamma
^2-4\alpha\beta)^{1/2},$ with $J=\cosh\Gamma+i\hbar\gamma\sinh\Gamma/\Gamma
=1/(1-i\hbar G).$
Having performed a straightforward calculation, we get the explicit
form of the density operator:

  \bea  \rho={\hbar\over \sqrt{\xi}}\exp\{{1\over 2}\ln{\xi\over \xi-i
\hbar\chi'}
-{1\over 2\hbar\sqrt{\xi-i\hbar\chi'+{1\over 4}\hbar^2}}\cosh^{-1}(1+{\hbar^2
\over 2(\xi-i\hbar\chi')})\nonumber\\
  \times  [ \phi(\hat q-<\hat q>)^2+\psi(\hat p-<\hat p>)^2-(\chi'+i{\hbar
\over 2})
[2(\hat q-<\hat q>)(\hat p-<\hat p>)-i\hbar] ] \},
      \eea
The density operator (29) is in a Gaussian form, as was expected from the
initial form of the Wigner distribution function. While the density operator is
expressed in terms of operators $\hat q$ and $\hat p,$ the Wigner distribution
is a function of real variables $q$ and $p.$ When time $t$ goes to infinity,
the density operator approaches to

\bea  \rho(\infty)={\hbar\over{\sqrt{\sigma-{1\over 4}\hbar^2}}}
\exp\{-{1\over
2\hbar\sqrt{\sigma}}\ln{2\sqrt{\sigma}+\hbar\over 2\sqrt{\sigma}-\hbar}
[\sigma_{pp}(\infty)\hat q^2+\sigma_{qq}(\infty)\hat p^2-\sigma_{pq}(\infty)
(\hat q\hat p+\hat p\hat q)]\},
    \eea
where $\sigma=\sigma_{pp}(\infty)\sigma_{qq}(\infty)-\sigma^2_{pq}(\infty)$
and \cite{ss}:

\bea   \sigma_{qq}(\infty)={1\over 2(m\omega)^2\lambda(\lambda^2+\Omega^2)}
[(m\omega)^2(2\lambda(\lambda+\mu)+\omega^2)D_{qq}
  +\omega^2D_{pp}+2m\omega^2(\lambda+\mu)D_{pq}], \eea
\bea   \sigma_{pp}(\infty)={1\over 2\lambda(\lambda^2+\Omega^2)}[(m
\omega)^2
\omega^2D_{qq}+(2\lambda(\lambda-\mu)+\omega^2)D_{pp}-2m\omega^2(\lambda-
\mu)D_{pq}],    \eea
\bea   \sigma_{pq}(\infty)={1\over 2m\lambda(\lambda^2+\Omega^2)}[-
(\lambda+
\mu)(m\omega)^2D_{qq}+(\lambda-\mu)D_{pp}+2m(\lambda^2-\mu^2)D_{pq}].
\eea
In the particular case (9),
\bea  \sigma_{qq}(\infty)={\hbar\over 2m\omega}\coth{\hbar\omega\over 2kT},~
\sigma_{pp}(\infty)={\hbar m\omega\over 2}\coth{\hbar\omega\over 2kT},~
 \sigma_{pq}(\infty)=0    \eea
and the asymptotic state is a Gibbs state (8):
\bea  \rho_G(\infty)=2\sinh{\hbar\omega\over 2kT}\exp\{-{1\over kT}({1
\over 2m}
\hat p^2+{m\omega^2\over 2}\hat q^2)\}.\eea

\section{Entropy and effective temperature}

Because of the presence of the exponential form in the density operator, the
construction of the logarithmic density is straightforward.
In view of the relations (22)-(24), the expectation value of the logarithmic
density becomes
\bea  <\ln\rho>=\ln\hbar-{1\over 2}\ln(\delta-{\hbar^2\over 4})-{\sqrt{\delta}
\over
\hbar}\ln{2\sqrt{\sigma}+\hbar\over 2\sqrt{\sigma}-\hbar}.\eea
By putting $\hbar\nu=\sqrt{\delta}-\hbar/2,$
we finally get the entropy in a closed form,
\bea  S(t)=k[(\nu+1)\ln(\nu+1)-\nu\ln\nu].    \eea
Because of the identity $\delta=-{\displaystyle{\hbar\omega^2\over 4\Omega^2}}
B_w,$
the function $\nu$ takes explicitly
the form
\bea  \nu={\omega\over 2\Omega}\sqrt{-B_w}-{1\over 2},\eea
where
\bea B_w=\exp(-4\lambda t)(2{\mu\over\omega}{\rm Re}{d_1 a^*\over\Lambda}
-{\Omega^2\over\omega^2}+{\vert d_1\vert^2\over
\vert\Lambda\vert^2}-
{d_2^2\over\lambda^2}+2{d_2\over\lambda})\nonumber\\
-2\exp(-2\lambda t) [{\rm Re}
\left ( ({\mu\over\omega}{ d_1 a^*\over\Lambda}+{\vert d_1\vert^2
\over
\vert\Lambda\vert^2})\exp 2i\Omega t \right )-
{d_2^2\over\lambda^2}+{d_2\over\lambda} ]+{\vert d_1\vert^2\over
\vert\Lambda\vert^2}-
{d_2^2\over\lambda^2}.
\eea
It is worth noting that the entropy depends only upon the variance of the
Wigner distribution.
When time $t\to\infty,$ the function $\nu$ goes to $s=\omega(d_2^2/
\lambda^2-\vert d_1\vert^2/(\lambda^2+\Omega^2))^{1/2}/2\Omega-1/2$
and the entropy relaxes to its equilibrium value $S(\infty)=k[(s+1)\ln(s+1)-s
\ln s].$

A similar expression to (37) has been formerly obtained \cite{lw,rh} in the
framework of the quantum theory of oscillator relaxation.
It should also be noted that the expression (37) has the same form as the
entropy of a system of harmonic oscillators in thermal equilibrium. In the
later case $\nu$ represents, of course, the average of the number
operator \cite{a1}.
Eq. (37) together with the function $\nu$ defined by (38), (39) is the desired
entropy
for the system. While the formal expression (37) for the entropy has a
well-known
appearance, the explicit form of the function $\nu$ displays
clearly a
specific
feature of the present entropy.
We see that the time
dependence of the entropy is represented by the damping factors $\exp(-4
\lambda t),~ \exp(-2\lambda t) $ and also by the oscillating function $\exp
2i\Omega t.$ The complex
oscillating factor $\exp 2i\Omega t$ occurs also in the second moments, as
shown in \cite{a}.
This factor reduces, however, to a function of the frequency $\omega,$
expressly $\exp 2i\omega t$ for $\mu\to 0$ or if $\mu/\Omega\ll 1$ (i.e.
the frequency $\omega$ is very large as compared to $\mu).$

In the case of a thermal bath (8), (9), we may define a time-dependent
effective temperature \cite{lw} $T_e,$
by remarking that at infinity of time
the quantity $\nu$ goes, according to (2.22) in ref. \cite{a},
to the average thermal
phonon number
$<n>=(\exp(\hbar\omega/kT)-1)^{-1}.$
Thus $\nu$ may be
considered as the time variation of the thermal phonon number.
Accordingly we may put in this case
\bea  (\exp{\hbar\omega\over kT_e}-1)^{-1}=\nu.\eea
The function $\nu,$ which goes to $<n>$ as $t$ tends to infinity, vanishes at
$t=0.$ From (40) the effective temperature $T_e$ can be
extracted as
\bea  T_e(t)={\hbar\omega\over k[\ln(\nu+1)-\ln\nu]}.\eea
In terms of effective temperature we may say that the system at
time $t$ is found in thermal equilibrium at temperature $T_e.$
Then, in terms of the temperature, the entropy takes the form
\bea S={\hbar\omega\over T_e(\exp{\hbar\omega\over kT_e}-1)}-
k\ln[1-\exp(-{\hbar\omega\over kT_e})].\eea
This form of the entropy for the thermally excited oscillator state
is well-known.
The
effective temperature approaches thermal equilibrium with the bath,
$T_e\to T,$ as
the value of $t$ increases.

\section{Concluding remarks}

Recently we assist to a revival of interest in quantum brownian motion as a
paradigm of quantum open systems. There are many motivations. The possibility
of preparing systems in macroscopic quantum states led to the problems of
dissipation in tunneling and of loss of quantum coherence (decoherence). These
problems are intimately related to the issue of quantum-to-classical
transition.
All of them point the necessity of a better understanding of open quantum
systems
and all requires the extension of the model of quantum brownian motion.
Our results allow such extensions.
The Lindblad theory provides a selfconsistent
treatment of damping as a possible extension of quantum mechanics to open
systems. In the present paper we have studied the one-dimensional harmonic
oscillator with dissipation within the framework of this theory.
We have first obtained the density operator from the master and
Fokker-Planck equations.
The density operator
in a Gaussian form is a function of the position and momentum operators
in addition to several time dependent
factors. The explicit form of the density operator has been subsequently used
to calculate the entropy and the effective temperature.
The temporal behaviour of these quantities
displays how they relax
to the equilibrium value.
In a future work we plan to use the entropy of the system in order to quantify
the extent of decoherence in the course of the system-environment interaction.


\end{document}